\documentclass[a4paper,aps,prd,10pt,preprintnumbers,showpacs,twocolumn,superscriptaddress,nofootinbib,amsmath,amssymb,floatfix]{revtex4-1}\def\PRDstyle#1{#1}\def\JCAPstyle#1{}\def\Abstract#1{\begin{abstract}#1\end{abstract}}

\usepackage{graphicx}
\usepackage[utf8]{inputenc}
\usepackage[T1]{fontenc}
\usepackage{cmap}

\def\imo{i}

\def\K{{\cal K}}
\DeclareMathAlphabet{\pazocal}{OMS}{zplm}{m}{n}

\begin{document}
\title{Black holes in galactic centers: quasinormal ringing, grey-body factors and Unruh temperature}
\JCAPstyle{
\author[\star\dagger]{R. A. Konoplya,}\emailAdd{roman.konoplya@gmail.com}
\affiliation[\star]{Research Centre for Theoretical Physics and Astrophysics,\\ Institute of Physics, Silesian University in Opava,\\ Bezručovo náměstí 13, CZ-74601 Opava, Czech Republic}
\affiliation[\dagger]{Peoples Friendship University of Russia (RUDN University),\\ 6 Miklukho-Maklaya Street, Moscow 117198, Russian Federation}
}
\PRDstyle{
\author{R. A. Konoplya}\email{roman.konoplya@gmail.com}
\affiliation{Research Centre for Theoretical Physics and Astrophysics, Institute of Physics, Silesian University in Opava, Bezručovo náměstí 13, CZ-74601 Opava, Czech Republic}
\affiliation{Peoples Friendship University of Russia (RUDN University), 6 Miklukho-Maklaya Street, Moscow 117198, Russian Federation}
}

\Abstract{
Recently, Cardoso et al. \cite{Cardoso:2021wlq} found an exact solution describing the black hole immersed in a galactic-like distribution of matter.
There, the properties of gravitational radiation were studied. Here we continue analysis of properties of this geometry via consideration of the electromagnetic radiation. We calculate quasinormal modes, asymptotic tails and grey-body factors for electromagnetic radiation. In addition, we discuss the Unruh temperature for this spacetime. Estimations made in the regime which is best fitting the galaxies behavior show that influence of the environment on classical and quantum radiation around such black holes must be relatively small.}

\maketitle

\section{Introduction}

Perturbations, quasinormal modes, radiation and other effects  around black holes immersed in astrophysical environment have been actively studied in the literature (see for instance \cite{Visser:1992qh,Bamber:2021knr,Leung:1997was,Konoplya:2019sns,Konoplya:2018yrp,Macedo:2015ikq} and references therein). In this context a recent work by Cardoso et. al.  \cite{Cardoso:2021wlq} suggested an interesting exact solution which describes a black hole immersed in the distribution of matter (with anisotropic pressure) which
meets observational data for galaxies. This way such a solution can serve as a model for a black hole immersed in the center of a galaxy. The gravitational perturbations, as well as basic aspects of particle motion, were considered in \cite{Cardoso:2021wlq}. Here we will study propagation of the electromagnetic perturbations in the background found in \cite{Cardoso:2021wlq}. We will study all stages of decay of electromagnetic perturbations, which includes not only quasinormal modes, but also the late time tails.
The scattering properties will be studied via calculation of the grey-body factors. In addition, we will consider the Unruh temperature for a given spacetime, which is a characteristic of the surface gravity at a given point in space within the galactic halo.

Our paper is organized as follows. In Sec. II we briefly describe the metric under consideration. Sec. III deduces the master wave equation for the electromagnetic field. Sec. IV discusses the methods used for analysis of quasinormal modes and late time tails and include the numerical data for quasinormal modes and asymptotic tails. Sec. IV is devoted to calculations of grey-body factors. Sec. V discusses the Unruh temperature of the spacetime as a function of the radial coordinate.

\section{The geometry of a galactic black hole with Hernquist distribution of matter}

In \cite{Cardoso:2021wlq}  the Hernquist-type density distribution~\cite{Hernquist:1990be} was used for modelling  the S\'ersic profiles that are observed in bulges and elliptical galaxies:
\begin{equation}
\rho(r)=\frac{Ma_0}{2\pi\,r(r + a_0)^3}\,,\label{eq_rho:hernquist}
\end{equation}
where $M$ is the total mass of the ``halo'' and $a_0$ is a typical lengthscale of the galaxy under consideration.
The metric has the following general form:
\begin{equation}
ds^2=-f(r) dt^2+\frac{dr^2}{1-2m(r)/r}+r^2d\Omega^2\,,
\label{eq:Spherical}
\end{equation}

The matter distribution suggested by the Hernquist profile~\eqref{eq_rho:hernquist} is compatible with the following (though, not unique) choice of the mass function:
\begin{equation}
m(r)=M_{\rm BH}+\frac{M r^2}{(a_0+r)^2}\left(1-\frac{2M_{\rm BH}}{r}\right)^2\,,
\end{equation}
which corresponds to the asymptotic flatness ($f\to 1$ at large $r$).
At small distances the above profile describes a source of mass $M_{\rm BH}$ which goes over to the Hernquist distribution~\eqref{eq_rho:hernquist} at large scales.
Using the Hernquist distribution as that, generating the source term for the Einstein equations, the metric function $f(r)$ has been obtained in \cite{Cardoso:2021wlq}  in the form
\begin{eqnarray}
f(r)&=&\left(1-\frac{2M_{\rm BH}}{r}\right)e^\Upsilon\,,\label{eq_fhairy}\\
\Upsilon&=&-\pi\sqrt{\frac{M}{\xi}}+2\sqrt{\frac{M}{\xi}}\arctan{\frac{r+a_0-M}{\sqrt{M\xi}}}\,,\\
\xi&=&2a_0-M+4M_{\rm BH}\,.
\end{eqnarray}
Observation of galaxies' corresponds to the regime $$a_0 \gtrsim 10^{4}M$$. The matter density is
\begin{equation}
4\pi \rho(r)=\frac{m'(r)}{r^2}=\frac{2M(a_0+2M_{\rm BH})(1-2M_{\rm BH}/r)}{r(r+a_0)^3}\,.\label{eq:hernquist_GR}
\end{equation}
At large distances and for $a_0\gg M_{\rm BH}$, this density profile is the Hernquist one \cite{Hernquist:1990be}.
The event horizon is located then at $r=2M_{\rm BH}$. The ADM mass  is $M+M_{\rm BH}$.

We will mainly be interested in the regime,
$$ M_{BH} \ll M \ll a_{0},$$
which is astrophysically relevant.

\section{Master wave equations}\label{sec:waveequations}
The general covariant equations for the  electromagnetic field $A_\mu$  have the form:
\begin{equation}\label{perturbeqs}
\frac{1}{\sqrt{-g}}\partial_{\mu} \left(F_{\rho\sigma}g^{\rho \nu}g^{\sigma \mu}\sqrt{-g}\right)=0\,
\end{equation}
where $F_{\mu\nu}=\partial_\mu A_\nu-\partial_\nu A_\mu$.
After the separation of variables Eqs.~(\ref{perturbeqs}) can be reduced to the following Schrödinger-like form (see, for instance, \cite{Konoplya:2011qq,Berti:2009kk,Kokkotas:1999bd})
\begin{equation}\label{wave-equation}
\dfrac{\partial^2 \Psi}{\partial t^2}-\dfrac{\partial^2 \Psi}{\partial r_*^2}=V(r)\Psi,
\end{equation}
where the ``tortoise coordinate'' $r_*$ is defined by the relation
\begin{equation}
dr_*=\frac{dr}{\sqrt{f(r) (1- (2 m(r)/r))}}.
\end{equation}

The effective potentials for the electromagnetic field is
\begin{equation}\label{potentialScalar}
V(r)=f(r) \frac{\ell(\ell+1)}{r^2},
\end{equation}
where $\ell=1, 2, \ldots$ are the multipole numbers. In the astrophysically motivated range of parameters the effective potentials have the form of the positive definite potential barriers.

\section{Quasinormal modes and late time tails}\label{sec:QNMs}
\begin{table*}
\PRDstyle{\begin{tabular}{p{1cm}c@{\hspace{0.5cm}}c@{\hspace{0.5cm}}c@{\hspace{0.5cm}}c@{\hspace{0.5cm}}c@{\extracolsep{\fill}}}}
\JCAPstyle{\begin{tabular}{p{4em}cccc}}
\hline
$a_0/M_{BH}$ & $M=10 M_{BH}$ & $M=20 M_{BH}$ &  $M=30 M_{BH}$  & $M=100 M_{BH}$  & $M=200 M_{BH}$  \\
\hline
$2 \cdot 10^5$   & 0.496484 - 0.184984 i & 0.496459 - 0.184975 i & 0.496435 - 0.184966 i &            0.496261 - 0.184901 i  & 0.496013 - 0.184808 i\\
$10^5$   & 0.496459 - 0.184975 i  & 0.496410 - 0.184956 i & 0.496360 - 0.184938 i &            0.496013 - 0.184808 i  & 0.495516 - 0.184624 i\\
$2 \cdot 10^{4}$   & 0.496261 - 0.184901 i  & 0.496013 - 0.184804 i & 0.495765 - 0.184716 i  & 0.494029 - 0.184069 i  & 0.491553 - 0.183147 i\\
$10^{4} $   & 0.496013 - 0.184808 i  & 0.495517 - 0.184624 i & 0.495021 - 0.184439 i  &        0.491554 - 0.183146 i  & 0.486615 - 0.181306 i\\
$2000 $    & 0.494033 - 0.184069 i  & 0.491560 - 0.183146 i & 0.489092 - 0.182225 i  &         0.471932 - 0.175821 i  & 0.447778 - 0.166808 i\\
$1000 $    & 0.491568 - 0.183146 i  & 0.486644 - 0.181305 i & 0.481736 - 0.179471 i  &         0.447853 - 0.166808 i  & 0.400938 - 0.149283 i\\
$400 $     & 0.484247 - 0.180386 i  & 0.472086 - 0.175820 i & 0.460030 - 0.171293 i  &         0.378678 - 0.140793 i  & 0.272652 - 0.101157 i \\
$200 $     & 0.472278 - 0.175819 i  & 0.448451 - 0.166809 i & 0.425040 - 0.157969 i  &         0.274053 - 0.101254 i  & 0.106502 - 0.039012 i \\
$100 $     & 0.449182 - 0.166818 i  & 0.403434 - 0.149339 i & 0.359376 - 0.132591 i  &         0.109785 - 0.039542 i  & -- \\
\hline
\end{tabular}
\caption{Fundamental ($n=0$, $\ell=1$) quasinormal mode of the electromagnetic field as a function of $a_{0}$ calculated by the higher order WKB approach. The corresponding Schwarzschild mode is $\omega = 0.496527 - 0.184975 i$; $M_{BH}=1/2$. The quasinormal frequency for $a_{0}=100$, $M= 200 M_{BH}$ is omitted, because this regime does not describe a meaningful configuration.}\label{tabl:em1}
\end{table*}
\begin{table*}
\PRDstyle{\begin{tabular}{p{1cm}c@{\hspace{0.5cm}}c@{\hspace{0.5cm}}c@{\hspace{0.5cm}}c@{\hspace{0.5cm}}c@{\extracolsep{\fill}}}}
\JCAPstyle{\begin{tabular}{p{3em}cc}}
\hline
$a_0/M_{BH}$ & $M=10 M_{BH}$ & $M=20 M_{BH}$ & $M=30 M_{BH}$  & $M=100 M_{BH}$ & $M=200 M_{BH}$\\
\hline
$2 \cdot 10^5$   & 0.915146 - 0.190001 i & 0.915100 - 0.189991 i& 0.915054 - 0.189982 i  &  0.914734 - 0.189916 i   &  0.914277 - 0.189821 i \\
$10^5$   & 0.915100 - 0.189992 i & 0.915009 - 0.189973 i& 0.914917 - 0.189954 i  &               0.914277 - 0.189821 i   &  0.913362 - 0.189631 i \\
$2 \cdot 10^{4} $   & 0.914734 - 0.189916 i & 0.914277 - 0.189821 i  & 0.913819 - 0.189726 i  &  0.910620 - 0.189061 i   &  0.906056 - 0.188114 i \\
$10^{4} $   & 0.914277 - 0.189821 i & 0.913362 - 0.189631 i & 0.912448 - 0.189441 i &             0.906058 - 0.188114 i  &  0.896955 - 0.186223 i \\
$2000$    & 0.910626 - 0.189061 i & 0.906069 - 0.188114 i  & 0.901519 - 0.187168 i  &             0.869885 - 0.180590 i  &  0.825361 - 0.171334 i \\
$1000$    & 0.906083 - 0.188114 i & 0.897004 - 0.186223 i  & 0.887957 - 0.184339 i  &             0.825493 - 0.171334 i  &  0.739006 - 0.153336 i \\
$400$     & 0.89258 - 0.1852804 i & 0.870156 - 0.180591 i & 0.847924 - 0.175943 i  &              0.697927 - 0.144623 i  &  0.502461 - 0.103916 i \\
$200$     & 0.870491 - 0.180593 i & 0.826537 - 0.171344 i& 0.783355 - 0.162268 i  &               0.504938 - 0.104032 i  &  0.196153 - 0.040095 i \\
$100$     & 0.827812 - 0.171363 i & 0.743373 - 0.153426 i & 0.662079 - 0.136237 i  &              0.202044 - 0.040669 i  &  -- \\
\hline
\end{tabular}
\caption{Fundamental ($n=0$, $\ell=2$) quasinormal mode of the electromagnetic field as a function of $a_{0}$ calculated by the higher order WKB approach. The corresponding Schwarzschild mode is $\omega = 0.915191-0.190009 i$; $M_{BH}=1/2$. The quasinormal frequency for $a_{0}=100$, $M= 200 M_{BH}$ is omitted, because this regime does not describe a meaningful configuration.}\label{tabl:em2}
\end{table*}

Quasinormal modes $\omega_{n}$ are frequencies corresponding to solutions of the master wave equation (\ref{wave-equation}) with the requirement of the purely outgoing waves at  infinities and at the event horizon:
\begin{equation}
\Psi \propto e^{-\imo \omega t \pm \imo \omega r_*}, \quad r_* \to \pm \infty.
\end{equation}

\begin{figure*}
\resizebox{\linewidth}{!}{\includegraphics{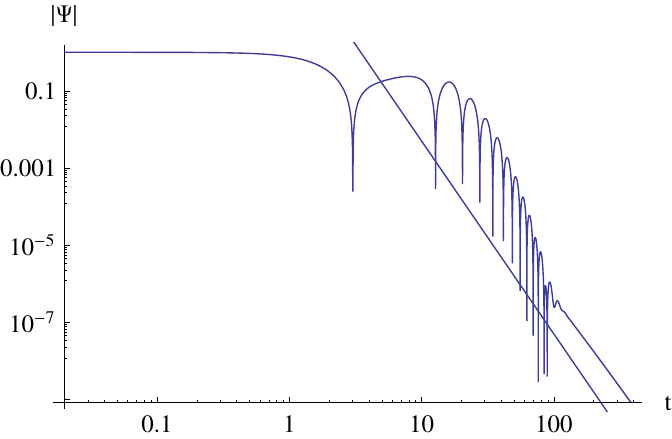}\includegraphics{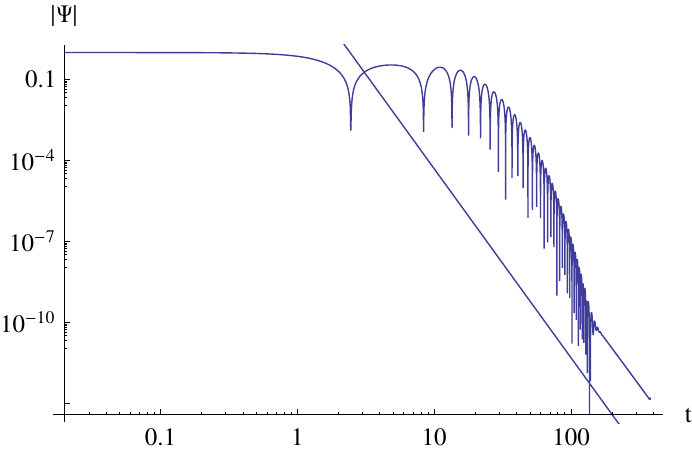}}
\caption{Logarithmic plots of late time tails for the electromagnetic radiation at $\ell=1$ (left) and $\ell=2$ right. The decay law is $|\Psi| \sim t^{-2 \ell -3}$.}\label{fig:Tails}
\end{figure*}

In order to find dominant quasinormal modes we will use the two methods:  the time-domain integration method \cite{Gundlach:1993tp} and the semi-analytic WKB method \cite{Schutz:1985zz,Iyer:1986np,Konoplya:2003ii,Matyjasek:2017psv}.

In the time domain, we can integrate the wavelike equation (\ref{wave-equation}) in terms of the light-cone variables $u=t-r_*$ and $v=t+r_*$.
We will apply the discretization scheme proposed in \cite{Gundlach:1993tp},
\begin{eqnarray}\label{Discretization}
\Psi\left(N\right)&=&\Psi\left(W\right)+\Psi\left(E\right)-\Psi\left(S\right)\PRDstyle{\\\nonumber&&}
-\Delta^2V\left(S\right)\frac{\Psi\left(W\right)+\Psi\left(E\right)}{4}+{\cal O}\left(\Delta^4\right)\,,
\end{eqnarray}
where the following notations for the points were used:
$N\equiv\left(u+\Delta,v+\Delta\right)$, $W\equiv\left(u+\Delta,v\right)$, $E\equiv\left(u,v+\Delta\right)$, and $S\equiv\left(u,v\right)$. The Gaussian initial data are imposed on the two null surfaces, $u=u_0$ and $v=v_0$. Then, the dominant quasinormal frequencies can be extracted from the time-domain profiles with the help of the Prony method \cite{Prony}.

In the frequency domain we will  use the WKB method of Will and Schutz \cite{Schutz:1985zz}, which was extended to higher orders in \cite{Iyer:1986np,Konoplya:2003ii,Matyjasek:2017psv} and achieved even higher accuracy via using of the Padé approximants \cite{Matyjasek:2017psv,Hatsuda:2019eoj}.
The higher-order WKB formula has the following form \cite{Konoplya:2019hlu},
\begin{eqnarray}
 \omega^2&=&V_0+A_2(\K^2)+A_4(\K^2)+A_6(\K^2)+\ldots \\\nonumber
&-& \imo \K\sqrt{-2V_2}\left(1+A_3(\K^2)+A_5(\K^2)+A_7(\K^2)+\ldots\right),
\end{eqnarray}
where $\K$ is half-integer. The corrections $A_k(\K^2)$ to the eikonal formula are of the order $k$  and polynomials in $\K^2$ with rational coefficients.  The corrections $A_k(\K^2)$ depend on the values of higher derivatives of the potential $V(r)$ in its maximum. In order to increase accuracy of the WKB formula, we will follow Matyjasek and Opala \cite{Matyjasek:2017psv} and use the Padé approximants. Here we will use the sixth order WKB method with $\tilde{m} =5$ (where $\tilde{m}$ is defined in \cite{Matyjasek:2017psv,Konoplya:2019hlu}), because this choice provides the best accuracy in the Schwarzschild limit and a good concordance with the time-domain integration.

Since both the WKB and time-domain integration methods are extensively used and discussed in the literature (see, for example, reviews \cite{Konoplya:2019hlu,Konoplya:2011qq}), we will not describe them in this paper, but will simply show that both methods are in a good agreement in the common parametric range of applicability.

From Tables (\ref{tabl:em1}, \ref{tabl:em2}) we see that the larger is the ratio $a_{0}/M$ at a fixed mass of the black hole, that is, the more rarefied is the environment, the closer are the  quasinormal frequencies to their Schwarzschild values. For the best fitting of the galaxies' behavior, this occurs at $a_{0} \gtrsim 10^4 M$ in which case the difference with the Schwarzschild values of quasinormal modes is only a small fraction of one percent. This agrees qualitatively with the relatively soft changes of the time-domain profile of quasinormal ringing in fig. 1 of \cite{Cardoso:2021wlq} for gravitational perturbations.  When $a_{0}/M$ is decreased, both real oscillation frequencies and damping rates are decreased, while their ratio $Re(\omega)/Im(\omega)$, characterising the quality factor of the oscillations is slightly increasing, which, rather counter-intuitively, makes the black hole a better oscillator when the galactic environment is taken into consideration.  This relatively small effect is difficult to predict from the form of the metric and it cannot be excluded that it is not general and appropriate to the particular distribution of matter under consideration.

In the eikonal regime $\ell \gg 1$ and taking into account that both $a_{0}$ and $M$ must be much larger than $M_{BH}$ and $a_{0} \gg M$ (see, for example \cite{Churilova:2019jqx} and references therein), we can use the first order WKB formula and find that the quasinormal frequencies obey the following relations:
\begin{equation}
Re (\omega) \approx \frac{1}{3 \sqrt{3} M_{BH}} \left(\ell + \frac{1}{2}\right) \sqrt{1- \frac{2 M}{a_{0}}} + {\cal O}\left(\ell^{-1}\right),
\end{equation}
\begin{equation}
Im (\omega) \approx - \frac{1}{3 \sqrt{3} M_{BH}} \left(n + \frac{1}{2}\right) \left(1- \frac{ M}{a_{0}}\right) + {\cal O}\left(\ell^{-1}\right).
\end{equation}
In the limit $M \rightarrow 0$, or, alternatively $a_{0} \rightarrow \infty$, the above expressions are reduced to those for the Schwarzschild case. Notice, that usually there is a correspondence between the eikonal quasinormal modes and the frequency and instability timescale of the circular null geodesics of a spherically symmetric asymptotically flat or de Sitter black holes \cite{Cardoso:2008bp}. As it was shown in \cite{Konoplya:2017wot,Konoplya:2019hml} this correspondence is not guaranteed for the gravitational and other non-minimally coupled fields, but it always takes place for minimally coupled fields, which is our case here.

Usually the decay law at asymptotically late tails depends on spacetime behavior in the far region. For example, the effective dark matter  term in the conformal Weyl gravity \cite{Konoplya:2020fwg} drastically changes the late time behavior of a scalar field.
In our case the asymptotic tails for the perturbation are found via the time-domain integration in the regime $M_{BH} \ll M \ll a_{0}$ and shown on fig. \ref{fig:Tails}. There, one can see that the decay law at late times, up to the numerical accuracy, is the same as that for the Schwarzschild case \cite{Price:1972pw}
\begin{equation}
|\Psi| \sim t^{-2 \ell -3}.
\end{equation}

\begin{figure*}
\resizebox{\linewidth}{!}{\includegraphics{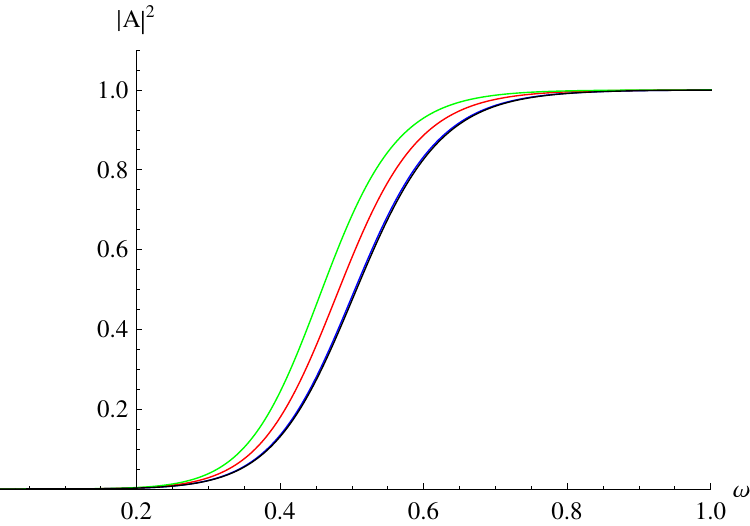}\includegraphics{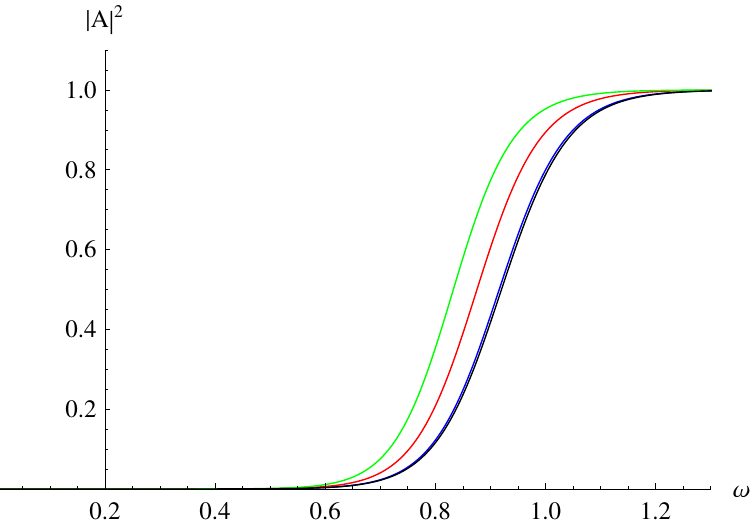}}
\caption{Grey body factors for the electromagnetic field $\ell =1$ (left) and $\ell=2$ (right), $M_{BH}=1/2$, $M= 10 M_{BH}$, $a_{0}= 2000 M$ (black, bottom), $a_{0}= 200 M$ (blue), $a_{0}= 20 M$ (red), $a_{0}= 10 M$ (green, top).}\label{fig:GreyBodyfromA0}
\end{figure*}

\begin{figure*}
\resizebox{\linewidth}{!}{\includegraphics{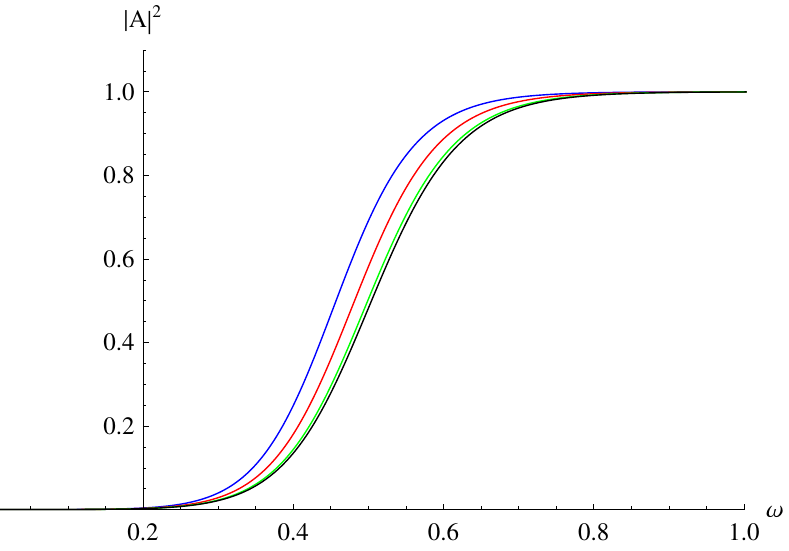}\includegraphics{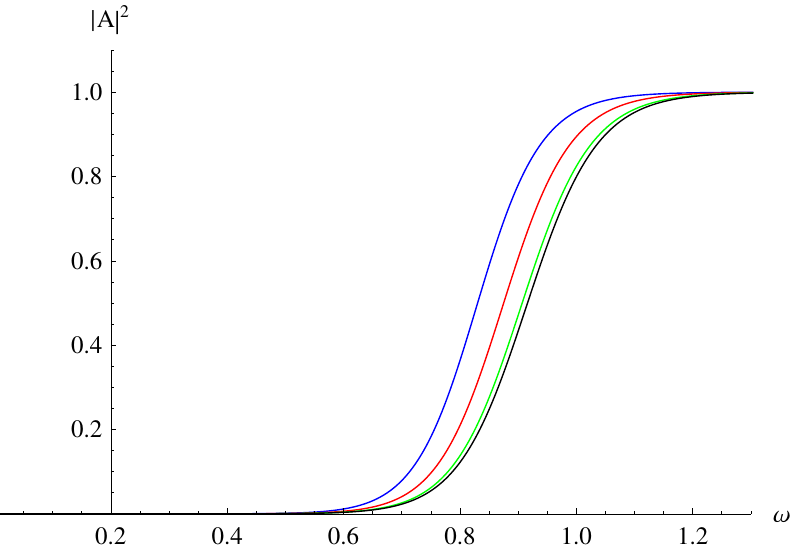}}
\caption{Grey body factors for the electromagnetic field $\ell =1$ (left) and $\ell=2$ (right), $M_{BH}=1/2$,  $a_{0}= 2000 M_{BH}$, $M= 10 M_{BH}$ (black, bottom), $M= 30 M_{BH}$ (green), $M= 100 0M_{BH}$ (red), $M= 200 M_{BH}$ (blue, top).}\label{fig:GreyBodyfromM}
\end{figure*}

\section{Grey-body factors}\label{sec:shadows}

Calculation of grey-body factors are important, first of all for estimation of the portion of the initial radiation in the vicinity of the event horizon which is reflected back to it by the potential barrier. For this one have to find the reflection and transmission coefficients or solve the so called \emph{scattering problem}, for which the boundary conditions are different from those required by quasinormal mode problem.

In the scattering problem we will consider the wave equation (\ref{wave-equation}) with the boundary conditions allowing for incoming waves from infinity. Owing to the symmetry of the scattering properties, this is identical to the scattering of a wave coming from the horizon. Then, the scattering boundary conditions for (\ref{wave-equation}) have the following form
\begin{equation}\label{BC}
\begin{array}{ccll}
    \Psi &=& e^{-i\omega r_*} + R e^{i\omega r_*},& r_* \rightarrow +\infty, \\
    \Psi &=& T e^{-i\omega r_*},& r_* \rightarrow -\infty, \\
\end{array}%
\end{equation}
where $R$ and $T$ are the reflection and transmission coefficients.
\par
The effective potential has the form of the potential barrier which monotonically decreases at both infinities, so that the WKB approach \cite{Schutz:1985zz,Iyer:1986np,Konoplya:2003ii} can be applied for finding $R$ and $T$. Since $\omega^2$ is real, the first order WKB values for $R$ and $T$ will be real \cite{Schutz:1985zz,Iyer:1986np,Konoplya:2003ii} and
\begin{equation}\label{1}
\left|T\right|^2 + \left|R\right|^2 = 1.
\end{equation}
Once the reflection coefficient is calculated, we can find the transmission coefficient for each multipole number $\ell$
\begin{equation}
\left|{\pazocal
A}_{\ell}\right|^2=1-\left|R_{\ell}\right|^2=\left|T_{\ell}\right|^2.
\end{equation}

From fig. \ref{fig:GreyBodyfromA0} we see that the smaller is the ratio $a_{0}/M$, the larger is the grey-body factors. In other words, the larger is the size of the halo under the same mass $M$, the bigger fraction of radiation will be scattered back to the black hole. The latter can be easily understood, because the effective potential at smaller $a_{0}/M$ becomes lower and, thereby, easier for transmission of radiation.  If the size of the halo is fixed, but its mass is varied, then the larger mass leads to larger  grey-body factors. However, as in the case of quasinormal modes, for $a_{0} \sim 10^4 M$ the estimation differ from the Schwarzschild case insignificantly.

\begin{figure*}
\resizebox{\linewidth}{!}{\includegraphics{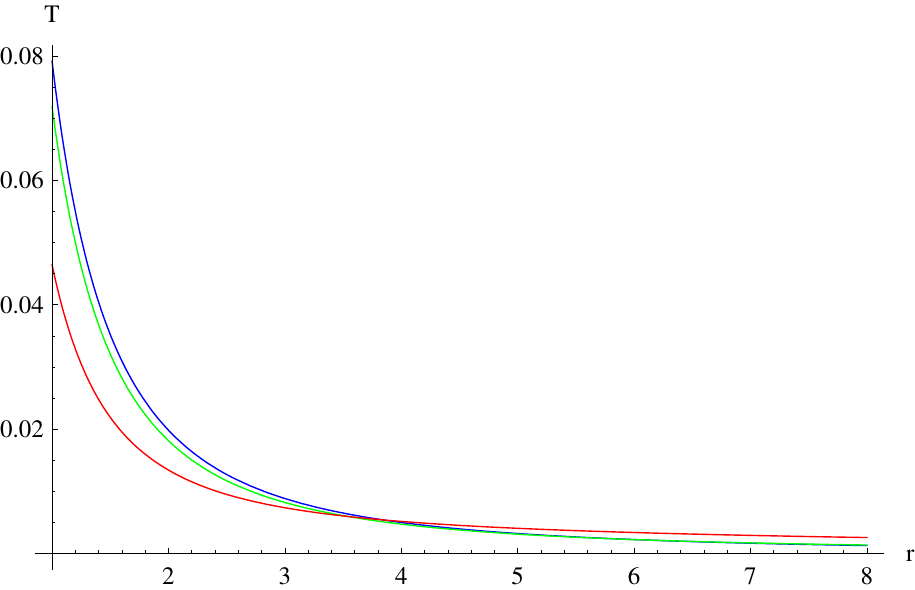}\includegraphics{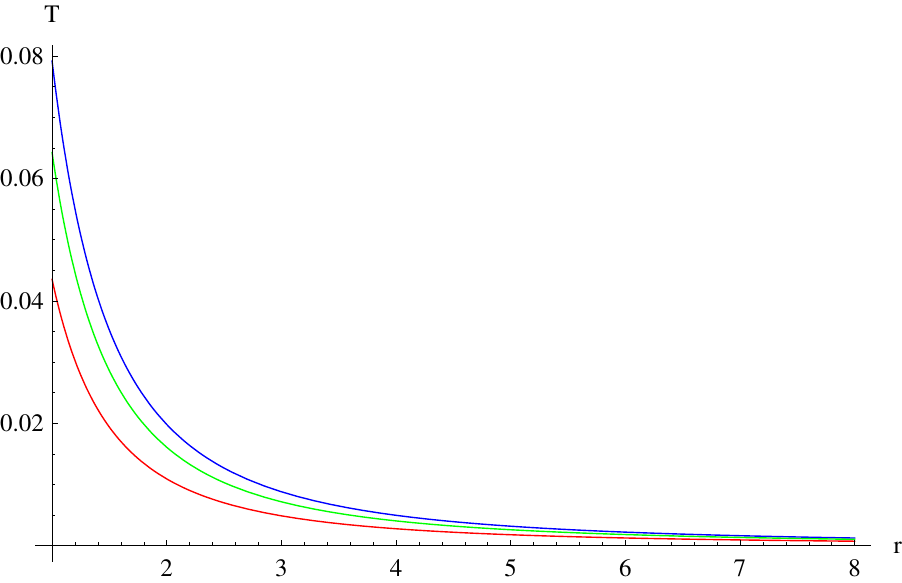}}
\caption{The Unruh temperature as a function of radial coordinate for various values of $a_{0}$ (left) and $M$ (right). Left panel: $M=10 M_{BH}$, $a_{0} =2000 M_{BH}$ (blue), $a_{0} =100 M_{BH}$ (green), $a_{0} =20 M_{BH}$ (red). Right panel: $a_{0} =1000$, $M= 10 M_{BH}$ (blue), $M= 400 M_{BH}$, $M= 1000 M_{BH}$ (red). }\label{fig:Unruh}
\end{figure*}

\section{Unruh temperature}

Here we would like to study the Unruh temperature \cite{Unruh1976} of the spacetime under consideration, because this temperature is a characteristic of the surface gravity (or acceleration) experienced by an observer at a given distance from the black hole. This characteristic was also studied within alternative approaches to gravitational theory \cite{Verlinde:2010hp}  (see also \cite{Konoplya:2010ak} and references therein).
In the general relativistic context one starts from a generalized form of the Newtonian potential
\begin{equation}
\phi = \frac{1}{2} \log (-g^{\alpha \beta }\xi_{\alpha} \xi_{\beta}),
\end{equation}
where $e^{\phi}$ is the red-shift factor that is supposed to be equal to unity at infinity ($\phi =0$ at $r= \infty$), if the space-time is asymptotically flat. The background metric is supposed to be some static solution which admits a global time-like Killing vector $\xi_{\alpha}$.

The acceleration is defined by the formula
\begin{equation}
a^{\alpha} = - g^{\alpha \beta} \nabla_{\beta} \phi,
\end{equation}
and the Unruh temperature on the shell of a fixed radius is given by the formula
\begin{equation}\label{T}
T = \frac{\hbar}{2 \pi} e^{\phi} n^{\alpha} \nabla_{\alpha} \phi,
\end{equation}
where $n_{\alpha}$ is a unit vector, that is normal to the Killing time-like vector $\xi_{\beta}$.
Using the equation (\ref{T}), the Unruh temperature can be written in the form
\begin{equation}\label{tt}
T = \frac{\hbar}{2 \pi} e^{\phi} \sqrt{g^{\alpha \beta} \phi_{,\alpha} \phi_{,\beta}} = \frac{\hbar}{4 \pi} \frac{f'(r)}{\sqrt{f(r) \left(1- \frac{2 m(r)}{r}\right)}}
\end{equation}

From fig. \ref{fig:Unruh} we see that larger values of mass of the halo $M$ result in a smaller surface gravity, while for larger ratio of $a_{0}/M$ this is not so: at smaller distance, large values of $a_{0}/M$ (i.e. more rarified environment) corresponds to larger surface gravity while in the far region, on the contrary, larger values of $a_{0}/M$ correspond to smaller surface gravity. This can be explained if we look at the above Unruh temperature in the regime of large $a_{0}$

\begin{widetext}
\begin{equation}\nonumber
T \approx \frac{M}{\pi  a_0^2} \left(\frac{1}{2}+ \frac{M_{BH}^2}{r^2}\right)  + \frac{M_{BH} \left(6 a_0(a_0 - M)+M (M-12 r)\right)}{12 \pi  a_0^2  r^2}.
\end{equation}

Here we used $\hbar =1$.
When $M$ is fixed in the latter expression, the second term becomes the subdominant one at sufficiently large $r$ (which, nevertheless, cannot be larger than $a_{0}$). Then, far from the black hole, the temperature is governed by the first term ($ \sim M (\pi a_0^2)^{-1}$), so that the larger $a_{0}$, the smaller is the surface gravity.

At the event horizon $r_{H}= 2 M_{BH}$ the Unruh temperature goes over into the Hawking temperature
$$
T_{H}= \frac{\sqrt{\exp \left(-\sqrt{\frac{M}{2 a_{0}-M+4 M_{BH}}} \left(\pi -2 \tan
   ^{-1}\left(\frac{a_{0}-M+2 M_{BH}}{\sqrt{M (2 a_{0}-M+4
   M_{BH})}}\right)\right)\right)}}{8 \pi  M_{BH}}
$$
\end{widetext}
When $a_{0}$ is large the Hawking temperature of the black hole horizon is
$$
T_H \approx \frac{a_{0}-M}{8 \pi  a_{0} M_{BH}},
$$
approaching the Schwarzschild value in the limit $M \rightarrow 0$. Unlike the Unruh temperature at a given $r$ as a characteristic of the acceleration or surface gravity, this expression for the Hawking temperature must be interpreted carefully, because, evidently, we cannot claim that the black hole is in a kind of thermal equilibrium with the whole galaxy. Instead, the thermodynamic equilibrium is established in a relatively small region near the black hole.

\section{Conclusions}\label{sec:conclusions}

Here we studied quasinormal modes, asymptotic late time tails and scattering properties of the electromagnetic radiation in the background of a black hole immersed in the Hernquist-type distribution of galactic matter. Estimations made in the range of parameters which corresponds to the best of experimental data for galaxies behavior show that influence of the environment upon radiation phenomena around such black holes must be relatively small for the observed electromagnetic quasinormal frequencies in the regime of the late time decay. However, there is a possibility that even such small differences can have significant cumulative effects due to the large number of orbits that an extreme mass ratio binary performs as was observed in \cite{Cardoso:2021wlq}.

\vspace{5mm}
\begin{acknowledgments}
I would like to acknowledge support of the grant 19-03950S of Czech Science Foundation (GAČR).
\end{acknowledgments}

\end{document}